\documentclass[prl,twocolumn,tightenlines,showpacs]{revtex4}%
\usepackage{amsfonts}
\usepackage{amsmath}
\usepackage{amssymb}
\usepackage{graphicx}

\begin{document}
\title{Macroscopic tunnel splittings in superconducting phase qubits}
\author{Philip R. Johnson}
\email[electronic address: ]{philipj@physics.umd.edu}
\author{William T. Parsons}
\author{Frederick W. Strauch}
\email[electronic address: ]{fstrauch@physics.umd.edu}
\author{J.R. Anderson}
\author{Alex J. Dragt}
\author{C.J. Lobb}
\author{F.C. Wellstood}
\affiliation{Department of Physics, University of Maryland, College Park, MD 20850}

\begin{abstract}
Prototype Josephson-junction based qubit coherence times are too short for
quantum computing. Recent experiments probing superconducting phase qubits
have revealed previously unseen fine splittings in the transition energy
spectra. These splittings have been attributed to new microscopic degrees of
freedom (microresonators), a previously unknown source of decoherence. We show
that macroscopic resonant tunneling in the extremely asymmetric double well
potential of the phase qubit can have observational consequences that are
strikingly similar to the observed data.

\end{abstract}
\date{\today}
\maketitle

Recent experiments by Simmonds \textit{et al.} \cite{Simmonds et al} and
Cooper \textit{et al. }\cite{Cooper et al} reveal previously unseen fine
splittings in the transition energy spectra of superconducting phase qubits.
These splittings are interpreted as resulting from coupling between the
circuit's collective dynamical variable (the superconducting phase describing
the coherent motion of a macroscopic number of Cooper pairs) and microscopic
two-level resonators, hereafter called \emph{microresonators}, within
Josephson tunnel junctions. Microresonators may be an important decoherence
mechanism \cite{Simmonds et al,Cooper et al,Decoherence mechanisms} for many
different superconducting qubit devices \cite{Phase qubit references,Charge
qubits,Flux qubits} with broader implications for Josephson junction physics
generally. Key questions remain however. Are all of the observed splittings
truly a microscopic property of junctions? Could they instead be a macroscopic
property of the particular circuit, or a combination of microscopic and
macroscopic phenomena?

In fact, macroscopic resonant tunneling (MRT) can produce spectral splittings
in multiwell systems by lifting degeneracies between the states of different
wells. These effects have been probed by Rouse \textit{et al.}, Friedman
\textit{et al}., and others \cite{MRT} in superconducting circuits involving
asymmetric double wells with a few left well states, and $\lesssim$ 10
right-well states. MRT effects have also been demonstrated by Crankshaw
\textit{et al.} \cite{Crankshaw et al} in three-junction flux qubits, another
system in which spurious splittings have been reported \cite{Plourde and
Bertet}. What is not obvious is that MRT effects can be important for systems
with extremely asymmetric double well potentials, like the rf SQUID phase
qubit \cite{Simmonds et al,Cooper et al}, that have hundreds or thousands of
right well states. In this Letter, we analyze the phase qubit in this limit
and show that MRT\ produces surprisingly complex observational consequences
that are strikingly similar to \emph{some} of the observed data \cite{Simmonds
et al,Cooper et al}. MRT is therefore a possible mechanism for fine splittings
in a phase qubit and requires further examination.

\begin{figure}[t]
\begin{center}
\includegraphics{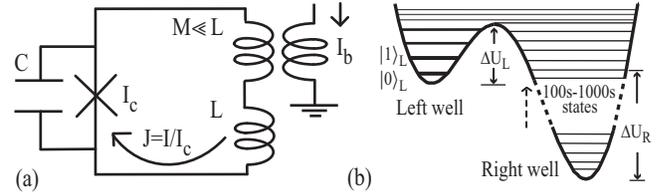}
\end{center}
\caption{(a) Circuit diagram for an rf SQUID phase qubit. (b) The device can
be tuned via an inductively coupled bias line to give an extremely asymmetric
double-well.}%
\end{figure}

Figure 1(a) shows the circuit schematic for an rf SQUID. The device is a
superconducting loop of inductance $L$ interrupted by a single Josephson
junction with capacitance $C$ and critical current $I_{c},$ inductively
coupled to a flux-bias line. The circuit Hamiltonian is%
\begin{equation}
H=4E_{C}p^{2}/\hbar^{2}+E_{J}\left(  \gamma^{2}/2\beta-\cos\gamma
-J\gamma\right)  ,\label{Hamiltonian}%
\end{equation}
where $\gamma$ is the gauge invariant phase difference across the junction,
$p=\hbar Q/2e$ is the momentum conjugate to $\gamma$ ($Q$ is the charge on the
plates of the capacitor), $\beta=2\pi I_{c}L/\Phi_{0}$ is the modulation
parameter ($\Phi_{0}=h/2e$ is the flux quantum), and $J=I/I_{c}$ is the
dimensionless current that is induced in the loop by the applied flux bias.
The charging energy $E_{C}=e^{2}/2C$ and Josephson energy $E_{J}=I_{c}\Phi
_{0}/2\pi$ determine the regime of superconducting qubit behavior; for a phase
qubit $E_{J}\gg E_{C}.$

\begin{figure*}[tb]
\begin{center}
\includegraphics{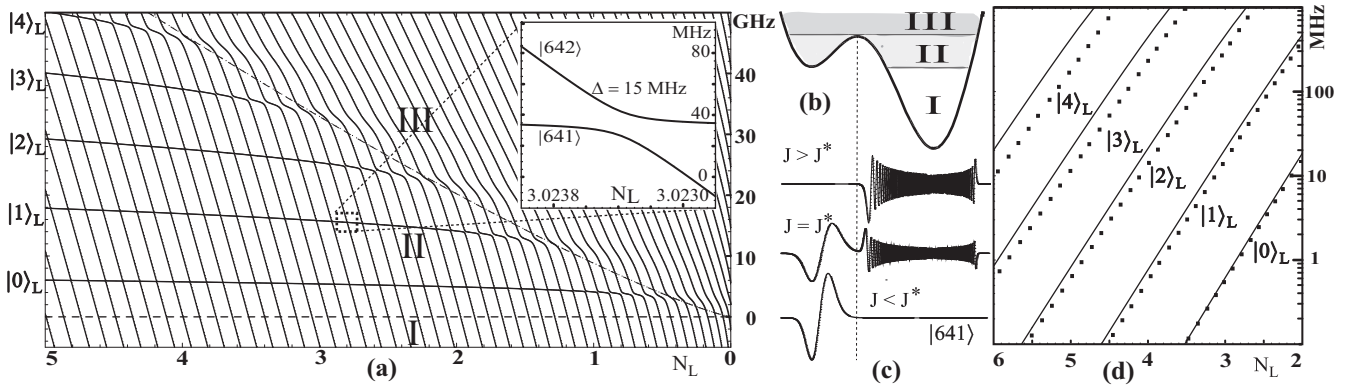}
\end{center}
\caption{(a) Numerically computed spectrum of phase qubit when $I_{c}=8.531$
$\mu$A, $C=1.2$ pF, and $L=168$ pH ($\beta=4.355$). Energies are plotted in
units of frequency. The inset shows the avoided crossing due to resonant
tunnel coupling between the left well state $\left\vert 1\right\rangle _{L}$
and a highly excited right well state. (b) The circuit parameters give an
asymmetric double well like that shown. (c) Wavefunctions of the $k=641$
eigenstate for bias values near the avoided crossing shown in the inset. (d)
Solid points are numerically computed sizes and locations of the splittings.
Solid lines are splitting sizes derived from WKB theory.}%
\end{figure*}

The shape of the circuit's potential energy function $U\left(  \gamma\right)
$ depends on $\beta$ and the bias $J.$ For $\beta\lesssim3\pi/2,$ it is
possible to bias the circuit so that the potential has the highly asymmetric
double-well shape shown in Fig.~1(b), tuned to give a shallow upper left well
with just a few left-localized states, denoted by $\left\vert n\right\rangle
_{L}$, and a deep right well with many right-localized states, denoted by
$\left\vert m\right\rangle _{R}$. Simmonds \textit{et al.} \cite{Simmonds et
al}--motivated by a number of attractive features including reduced
quasiparticle generation, tunable anharmoniticity of the left well potential,
inductive isolation from and reduced sensitivity to bias noise, and nice
read-out properties--have proposed using the rf SQUID with an extremely
asymmetric double well potential as a phase qubit \cite{Phase qubit
references}.

Making a cubic approximation to the left well, we derive the plasma frequency
for small oscillations%
\begin{equation}
\omega_{L}=\omega_{0}\left(  1-\beta^{-2}\right)  ^{1/8}\left[  2\left(
J^{\ast}-J\right)  \right]  ^{1/4},\label{w_L}%
\end{equation}
where $\omega_{0}=$ $\sqrt{8E_{c}E_{J}/\hbar^{2}},$ and%
\begin{equation}
J^{\ast}=\left(  1-\beta^{-2}\right)  ^{1/2}+\beta^{-1}\arccos\left(
-\beta^{-1}\right)  >1\label{J_star}%
\end{equation}
is the critical bias for which the left well vanishes. Note that the effective
critical current is $I^{\ast}=I_{c}J^{\ast}>I_{c}.$ The approximate number of
left-well states is%
\begin{equation}
N_{L}=\frac{\Delta U_{L}}{\hbar\omega_{L}}\simeq\frac{2^{3/4}}{3}\sqrt
{\frac{E_{J}}{E_{C}}}\left(  1-\beta^{-2}\right)  ^{-3/8}\left(  J^{\ast
}-J\right)  ^{5/4},\label{N_L}%
\end{equation}
where $\Delta U_{L}$ is the barrier height. The level spacing in the right
well is approximately $\hbar\omega_{R},$ where $\omega_{R}$ is the right well
plasma frequency, and the number of right well states is approximately
$N_{R}\simeq\Delta U_{R}/\hbar\omega_{R},$ where $\Delta U_{R}$ is the depth
of the right well.

Figure 2(a) shows the energy spectrum as $J$ is varied for $0\leq N_{L}\leq6$
and $C=1.2$ pF, $L=168$ pH, and $I_{c}=8.531$ $\mu$A, giving $\beta=4.355$,
$I^{\ast}=11.659$ $\mu$A, and $\omega_{L}/2\pi\sim10$ GHz. These are the
circuit parameters from \cite{Simmonds et al}, assuming that the critical
current quoted there is $I^{\ast}$. To obtain the energy spectrum we
diagonalize the Hamiltonian in Eq.~(\ref{Hamiltonian}) using a discrete
Fourier grid representation \cite{Fourier grid method}, thereby obtaining a
numerical solution for the eigenvalues $E_{k}\left(  J\right)  $ and
eigenstates $\left\vert k\left(  J\right)  \right\rangle $ of the full
double-well system versus the bias $J$. A harmonic approximation to the right
well yields approximately $500$ states below the left well; the full
calculation yields $N_{R}\simeq600-700$ states, depending on the bias
\cite{Footnote 2}.

In Fig.~2(a) we define the zero of energy to be at the bottom of the left
well. We note two different types of energy levels: horizontal ($H$) branches
and near vertical ($V$) branches. From our definition of zero energy,
eigenvalues corresponding to states mainly localized in the right well [region
I of Figs. 2(a) and (b)] fall with increasing $J,$ and are thus nearly
vertical. The energy levels in region III correspond to delocalized states
fully above the left well. The dashed line in Fig.~2(a) dividing regions II
and III indicates the energy at the top of the left-well barrier. In region
II, eigenstates whose energies lie along $H$ branches are primarily localized
in the left well ($H\sim L$). The number of left-well states at bias $J$ is
consistent with $N_{L}$ from Eq.~(\ref{N_L}). Eigenstates whose energies lie
along $V$ branches are primarily localized in the right well ($V\sim R$).
Their energies fall at essentially the rate of the falling right well. Note
that in Fig.~2(a) the \textit{density} of right-well states is comparable to
that of the left-well, despite $N_{R}\gg N_{L}.$

Every apparent intersection of an $H$ and $V$ energy level in Fig.~2(a) is an
avoided crossing (see inset). Degeneracies are lifted by resonant tunneling of
left-well states $\left\vert n\right\rangle _{L}$ and right-well states
$\left\vert m\right\rangle _{R}.$ Left of an avoided crossing between $k$ and
$k+1$ eigenstates we find that $\left\vert k\right\rangle \cong\left\vert
n\right\rangle _{L}$ and $\left\vert k+1\right\rangle \cong\left\vert
m\right\rangle _{R}$. Right of the crossing the states swap, becoming
$\left\vert k\right\rangle \cong\left\vert m\right\rangle _{R}$ and
$\left\vert k+1\right\rangle \simeq\left\vert n\right\rangle _{L}.$ At the
avoided crossing $\left\vert k\right\rangle \cong\left(  \left\vert
n\right\rangle _{L}+\left\vert m\right\rangle _{R}\right)  /\sqrt{2}$ and
$\left\vert k+1\right\rangle \cong\left(  \left\vert n\right\rangle
_{L}-\left\vert m\right\rangle _{R}\right)  /\sqrt{2}.$ Figure 2(c) shows the
wavefunctions for the $k=641$ eigenstate before, at, and after the splitting
shown in the inset in Fig.~2 (a). The distribution of splitting magnitudes
along the first five energy branches are plotted in Fig.~2(d) as solid points.
Gaps larger than 1 MHz are within the resolution of recent experiments. Along
each left-well energy branch the tunnel splittings are regularly spaced with
magnitudes that decrease exponentially with $N_{L}$. We have numerically
computed spectra for a variety of circuit parameters, including $I_{c}=2$
$\mu$A and $C=0.5$ pF which are comparable to those reported in \cite{Cooper
et al}. In each case the spectrum looks qualitatively similar to Fig.~2 (a).
We note that the predicted gap sizes are strikingly similar to those reported
in \cite{Simmonds et al,Cooper et al} ($\sim$ 1-100 MHz).

The complex collection of energy splittings has both direct and indirect
effects that should be taken into account when analyzing the experimental
data. Consider a double frequency microwave spectroscopic method, like that
used in \cite{Simmonds et al}. Microwaves of frequency $\omega_{01}$ are
applied to drive the $0\rightarrow1$ transition. Excitation of the $\left\vert
1\right\rangle _{L}$ state is detected with a measurement microwave pulse of
frequency $\omega_{13},$ which drives the $1\rightarrow3 $ transition. The
$\left\vert 3\right\rangle _{L}$ state's exponentially greater amplitude to be
found in the right well compared to the $\left\vert 0\right\rangle _{L}$ and
$\left\vert 1\right\rangle _{L}$ states allows an adjacent detection SQUID to
easily detect the change in the qubit's flux. This method directly probes
splittings along many of the energy branches shown in Fig.~2(d). Cooper
\textit{et al.} have introduced a new spectroscopic technique that can probe
deeper left wells where $N_{L}>4.$ This method applies a few-nanosecond
current pulse changing the bias so that $N_{L}\gtrsim2$ by briefly tilting the
potential adiabatically with respect to the left well period $T_{L}\equiv
2\pi/\omega_{L}\sim100$ ps \cite{Cooper et al}. Since the measurement pulse
moves left-well states to the right along horizontal ($H$) energy branches
[see Fig.~2(a)], read-out should be influenced by the exponentially larger
splittings present for smaller $N_{L}$. For example, the measurement pulse may
move a deep well state to one of the large splitting degeneracies near
$N_{L}\sim2,$ whose presence may produce a significant perturbation on
read-out fidelity. Thus the current pulse method is also sensitive to large
splittings along multiple energy branches.

MRT degeneracies also have very narrow bias value widths. For example, the
inset of Fig. 2(a) shows a splitting width of less than $0.1$ nA. The bias
widths become only smaller for splittings at larger $N_{L}.$ The horizontal
axis of Fig.~2(a) corresponds to more than $300$ nA. Typical experiments
sampling only a limited number of bias values likely probe only a subset of
the (many) MRT splittings. Changes in experimental conditions (e.g. bias drift
and noise, or temperature cycling) may generate surprisingly large shifts in
the observed splitting distributions if they result in a different subset of
sampled MRT\ degeneracies. These or other features could result in transition
spectra with a varying distribution of splitting sizes and bias-value
locations which, due to their complexity and variability in time, might appear
to have a microscopic origin. Such variations seem more consistent with a
model of microscopic critical current fluctuators, suggesting that both MRT
and microresonator effects are present \cite{Simmonds and Martinis}. If this
is the case, it is important to identify which observed splittings are due to
which mechanism.

Measured with sufficient resolution, the transition frequency avoided
crossings due to MRT should have distinctive characteristics. When driving
$0\rightarrow1$ transitions, a splitting in the $\left\vert 0\right\rangle
_{L}$ branch should produce crossings like that shown in Fig.~3(a), whereas a
splitting in the $\left\vert 1\right\rangle _{L}$ branch should produce
crossings like that shown in Fig.~3(b). The observed shapes may be strongly
dependent upon the experimental measurement technique. Bias noise could smear
out the splittings in the horizontal direction. For splittings in the lower
energy branch [Fig.~3(a)] this would leave a distinct frequency gap, givings
observed splittings horizontally smeared appearances like those observed in
\cite{Simmonds et al,Cooper et al}. In contrast, it is unclear if splittings
in the upper branch [Fig.~3(b)] are consistent with observation. Improved
experimental resolution that revealed these distinctive avoided-crossing
shapes would be compelling evidence for MRT. Han \textit{et al}. have explored
other complexities that arise when measuring systems that exhibit MRT
\cite{MRT}.

\begin{figure}[tb]
\begin{center}
\includegraphics{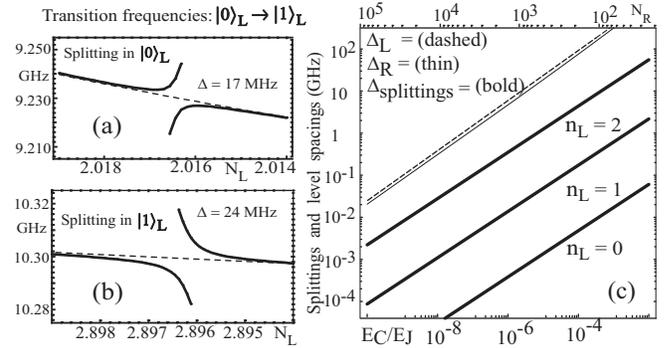}
\end{center}
\caption{(a) The distinctive shapes of avoided crossings in the measured
transition frequencies for splittings in the lower branch. (b) The avoided
crossing transition shape for splittings in the upper branch. (c) The figure
shows that $\Delta_{R}\approx\Delta_{L}$ over an extremely large range of
double-well circuit parameters. Bold lines show the splitting magnitudes
$\Delta$ along the $n_{L}=0,1,$ and $2$ left well energy branches, with
$\beta=4.5,$ $N_{L}=3,$ and $I_{c}=10\mu$A.}%
\end{figure}

We derive an analytic expression for the energy splitting between pair-wise
degenerate left and right states in an asymmetric double well in the WKB
approximation \cite{WKB methods}. This yields the splitting formula%
\begin{equation}
\Delta=\sqrt{\frac{2\Delta_{L}\Delta_{R}\left(  n_{L}+\frac{1}{2}\right)
^{n_{L}+1/2}\left(  m_{R}+\frac{1}{2}\right)  ^{m_{R}+1/2}}{\pi\ n_{L}%
!m_{R}!e^{n_{L}+m_{R}+1}}}e^{-S},\label{tunnel splitting formula}%
\end{equation}
where $S=\int_{\gamma_{1}}^{\gamma_{2}}\sqrt{2m\left[  E-V\left(
\gamma\right)  \right]  }d\gamma,$ $m=C\left(  \Phi_{0}/2\pi\right)  ^{2}$,
$\gamma_{1,2}$ are the classical turning points for the barrier given by
$V\left(  \gamma_{1,2}\right)  =E_{n_{L}}$, and $\Delta_{L}\simeq\hbar
\omega_{L}$, $\Delta_{R}\simeq\hbar\omega_{R}$ are the left and right well
level spacings at energy $E_{n_{L}}$. For deep right wells,
Eq.~(\ref{tunnel splitting formula}) becomes independent of $m_{R}.$ In this
limit, together with the cubic approximation accurate for shallow left wells,
the splittings are approximately%
\begin{equation}
\Delta\simeq\sqrt{\frac{2^{1/2}\Delta_{L}\Delta_{R}}{n_{L}!\pi^{3/2}}}\left(
432N_{L}\right)  ^{\left(  n_{L}+1/2\right)  /2}e^{-18N_{L}/5}%
.\label{cubic approx tunneling splitting formula}%
\end{equation}
For the right well level spacing we use the WKB estimate $\Delta_{R}=2\pi
\hbar/T_{cl}$ \cite{Landau and Lifshitz}, where $T_{cl}$ is the classical
period of right-well oscillations with energy $E_{n_{L}}.$ Splittings
calculated from Eq.~(\ref{cubic approx tunneling splitting formula}) are shown
as solid lines in Fig.~2(d). The agreement with the exact splittings (solid
points) is excellent for lower lying states, and surprisingly good for the
excited states. Note that the tunnel splitting formula in
Eq.~(\ref{cubic approx tunneling splitting formula}) predicts splittings
exponentially larger than continuum tunneling rates: $\Delta_{\text{splitting}%
}/\Gamma_{\text{tunneling}}\sim\exp\left(  18N_{L}/5\right)  ,$ making MRT
effects important even when continuum tunneling is negligible.

We have compared MRT splittings with
Eq.~(\ref{cubic approx tunneling splitting formula}) for a number of numerical
examples with $N_{R}\sim100-1000,$ but in principle one can fabricate circuits
with many thousands of right-well states. The WKB formula for the splittings
and level spacings allows analysis of circuit parameters for very deep right
wells where numerical treatment is impractical. Figure 3(c) shows $\hbar
\omega_{L}\simeq\Delta_{L}$ (dashed line) and $\Delta_{R}=2\pi\hbar/T_{cl}$
(thin-solid line) versus the ratio $E_{C}/E_{J}$ for $I_{c}=10\mu$A, $N_{L}=3
$, and $\beta=4.5$ just below the $\beta$ threshold where the potential
develops three wells. (For the circuit parameters in Fig.~2 and \cite{Cooper
et al}, $E_{C}/E_{J}\sim10^{-4}-10^{-6}$.) The value of $I_{c}$ determines the
frequency scale on the left of Fig.~3(c) but leaves the relative positions of
the plotted lines essentially unchanged. The top axis shows $N_{R}$ from the
harmonic oscillator approximation. Observe that, perhaps unexpectedly,
$\Delta_{R}\approx\Delta_{L}$ even for extremely asymmetric double wells. The
bold-solid lines show the WKB splitting $\Delta$ when $n_{L}=0,1,$ and $2$.
The validity condition for MRT $\Delta\ll\Delta_{R,L}$ is satisfied over a
large range of circuit parameters, and for $N_{R}\sim10^{5}$ and greater.

Dissipation suppresses resonant tunneling when $\Gamma_{R}\gtrsim\Delta_{R},$
where $\Gamma_{R}\simeq N_{R}\hbar/T_{1}$ is the width of excited right well
states, and $T_{1}$ is the dissipation time for $\left\vert 1\right\rangle
_{R}\rightarrow\left\vert 0\right\rangle _{R}$ \cite{Caldeira and
Leggett,Garg}. Using the WKB\ expression for $\Delta_{R},$ we find the
condition $N_{R}\lesssim\omega_{L}T_{1}$ for observing MRT. For a phase qubit
with $\omega_{L}/2\pi\sim10$ GHz and $T_{1}\sim10-100$ ns, resonant tunneling
should be detectable as long as $N_{R}\lesssim600-6000$ states. For the
circuit parameters in Fig.~2 $N_{R}\sim600-700$ and for those in \cite{Cooper
et al} $N_{R}\sim150-300,$ with a measured $T_{1}\simeq25$ ns. Thus, we do not
believe that dissipation will remove the effects of MRT. If the intrinsic
dissipation is actually much smaller so that $\Gamma\lesssim\Delta$
\cite{Garg}, it should be possible to observe coherent oscillations \cite{Long
T1}.

In conclusion, we show that significant MRT effects should be present for
extremely asymmetric double well phase qubits, and thus MRT should be taken
into account in the important effort to fully characterize microresonators or
other splittings mechanisms. Our analysis provides tools and can guide
experiments to help distinguish between three main possibilities: (1) Both MRT
and microresonators are present, (2) MRT effects explain all the observational
data, and (3) MRT is entirely absent. We believe that (1) is most likely;
however, due to the complexity of effects from MRT, further experiments and
detailed modeling are necessary to definitively rule out (2) and (3). Finally,
our Letter provides general tools for exploring the quantum mechanics of
extremely asymmetric double-well systems.

\begin{acknowledgments}
We thank R. Simmonds and J. Martinis for useful comments. This work was
supported by the NSA, the DCI Postdoctoral Research Program, the NSF QUBIC
program, DOE grant DE-FG02-96ER40949, and the University of Maryland's Center
for Superconductivity Research.
\end{acknowledgments}

\end{document}